\documentclass[floatfix,twocolumn,preprintnumbers,amsmath,amssymb]{revtex4}

\usepackage{graphicx}
\usepackage{dcolumn}
\usepackage{bm}
\begin{document}
\title{On the New Puzzling Results from MiniBooNE}
\author{Syed Afsar Abbas}
\email{afsarabbas@yahoo.com}
\author{Shakeb Ahmad}
\email{physics.sh@gmail.com}
\affiliation{Department of Physics, Aligarh Muslim University, Aligarh - 202002, India. }
\date{\today}
\begin{abstract}
We look into the recent puzzling results from MiniBooNE and contrast their results with that of NOMAD.
A picture which provides a consistent description of both is discusssed here. This also points to future
directions in neutrino studies. 
\end{abstract}
\maketitle

Recently MiniBooNE experiment has extracted the first measurment of the double differential cross section
for CCQE scattering of muon neutrinos on carbon and from this they obtained the single differential cross section and the absolute
cross section~\cite{a}. An effective axial mass of $M_A^{\text{eff}}=1.35\pm 0.17$ GeV, significantly higher than the 
historical world average value was extracted. Here we would like to offer an explanation of this unexpected
result.

Depending upon the available energies, upon the nature of the various particles interacting with nuclei, and upon 
the physical quantities under study, the nucleus offers a rich spectrum right from the quark-gluon to the
nucleonic and further to the cluster degrees of freedom. For example to understand the experimentally
determined 'hole' (i.e. a significant depression in central density) in $^{3}{\text H}$, $^{3}{\text He}$
and $^{4}{\text He}$ both the quark and the nucleonic degree of freedom manifest themselves simultaneously~\cite{b}.
For example for $A=3$ nuclei, the wave function that works well is
\begin{equation}
\Psi(^{3}{\text{He}})=a\psi(\text{ppn})+b\phi(\text{9q})
\end{equation}
where $\psi(\text {ppn})$ is significant for approximate distances $0.7\leq r\leq 1.8~\text{fm}$ (size of $^{3}{\text He}$) and
$\phi(\text{9q})$ for $r\leq 0.7~\text{fm}$ approximately where the three nucleons overlap strongly~\cite{b}. 

In an analogous manner we suggest that the degrees of freedom relevant for neutrino interacting with $^{12}{\text C}$ is
\begin{equation}
\Psi(^{12}{\text C})=a\psi(\text{6p,6n})+b\phi(\text{2t,2h})
\end{equation}
where triton $\text t\equiv~ ^{3}{\text H}$ and helion $\text h\equiv~ ^{3}{\text{He}}$.
The first term represents the standard shell model structure of $^{12}{\text C}$ as consisting of 
6 protons and 6 neutrons. The second term $\phi(\text{2t,2h})$ represents the clusters of $A=3$
kind i.e. $\text{2t+2h}$. Now $^{12}{\text C}$ has often been treated as made up of three alpha-clusters.
However for neutrino charge changing interactions, it shall play no role here.

To start with, let us accept the wave function (2) and derive the consequences. later we
shall show why the above structure should be physically acceptable and thus provide a consistent and
valid description of $^{12}{\text C}$ for the neutrino experiments. Now given Eqn.(2) for
$\Psi(^{12}{\text C})$ two kind of simultaneous knockout processes may ocuur
\begin{eqnarray}
\nu_\mu+n\rightarrow p+\mu^-\\
\nu_\mu+t\rightarrow h+\mu^-
\end{eqnarray}
The process (3) is represented by Fig.4 of ref.~\cite{a}. Here we draw another figure (Fig.1) representing
the process ocuuring from the reaction (4).
\begin{figure}[h]
\includegraphics[height=5cm,width=6cm]{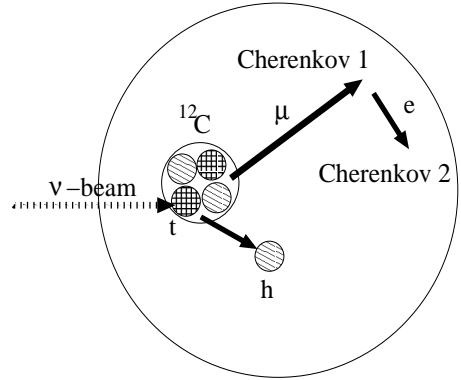}
\caption{Schematic illustration of the CCQE interaction of reaction (4) in addition to
Fig.4 (of ref~\cite{a}).}
\end{figure}

Now very often triton $\text t\equiv~ ^{3}{\text H}$ and helion $\text h\equiv~ ^{3}{\text{He}}$.
have been used as elementry in Elementry Particle Model(EPM) in $\nu_\mu/\mu$ interactions on these nuclei~\cite{c}
to provide a fruitful description of experimental data. We suggest similar perspective
of $^{12}{\text C}$ described as made up of 2t and 2h. Just as in Eq.(1) there are two kinds of independent
fermi gasses-that of nucleons $(ppn)$ and another one of quarks $(9q)$, here in Eq.(2)
there are two independent Fermi gas pictures relevant for $^{12}{\text C}$- the standard one
of the nucleons $(6p,6n)$ and the other one of the $A=3$ entities (called nusospin- see below).

Here as per our model the $\nu_\mu$ beam knocks out an $\text h(^{3}\text{He})$. As the MiniBooNE group 
is not observing the outgoing proton (or helion), what they are observing actually is the
cumulative effects due to these two independent channels. As all the models discussed by them~\cite{a}
correspond to $\psi(\text{6p,6n})$ term and modifications thereof, these models are actually completely
missing the contribution arising from the $\phi(\text{2t,2h})$ state.

Hence we suggest that the excess $\sim 30\%$ enhancement in cross section is due to the missing 
$\phi(\text{2t,2h})$ term in their analysis. Neutrino charged current terms are unique in that these are 
picking up the two special structures of $^{12}{\text C}$ uniquely. If we take $\psi(\text{6p,6n})$
and $\phi(\text{2t,2h})$ as two independent Fermi gas structure, then as in $\phi(\text{2t,2h})$
only A/3 cluster structures arise (i.e. 4 for $^{12}{\text C}$), there should be only 1/3 of structure 
associated with the $\phi(\text{2t,2h})$ term vis-a-vis the $\psi(\text{6p,6n})$ term.
Hence this is consistent with the fact that they obtain 30\% excess strength in their work~\cite{a}.

Interestingly our new picture here not only offers a clearcut explanation of the MiniBooNE data
but also explain the apparantly conflicting results of NOMAD~\cite{d}. The conflict between 
MiniBooNE and NOMAD data is very striking. NOMAD being 30\% lower than the MiniBooNE data and is 
fitted well by most nuclear physics models and without modifying the world-average value of $M_A$.

The resolution to this conflict within our model suggested here is that the NOMAD group spent a consistent amount
of efforts in identifying 1-track and 2-track effects. The 2-track events were those where the
knocked out proton in $\nu_\mu+n\rightarrow p+\mu^-$ reactions was specifically identified
and studied. Apparently they have done a good job of it and hence as per our model Eq.(3)
only $\psi(\text{6p,6n})$ has contributed in their study. That is, the contribution due to
$\nu_\mu+t\rightarrow h+\mu^-$ has been effectively filtered out by the NOMAD experiment.
Hence it is 30\% lower than the MiniBooNE experiment. Simultaneously undertanding of the conflicting results of the 
MiniBooNE~\cite{a} and NOMAD~\cite{d} experimental results should be taken as a proof of consistency
and validity of our model.

Now a few words on to why and how does the $^{12}{\text C}$ structure as given in Eq.(2) arise. In general
we take $^{12}{\text C}$ as made up of 6 protons and 6 neutrons in Shell Model. Sometimes
$^{12}{\text C}$ as made up of 3$\alpha$ is also invoked (but not relevant for charged current
$\nu$-interactions). Here we may treat $^{12}{\text C}$ as made
up of two $^{6}\text{Li}$ clusters. Next $^{6}\text{Li}$ may have h-t structure in the ground state.
Infact there are strong experimental evidences supporting $^{6}\text{Li}\equiv \text{t+h}$
(\cite{e}-\cite{f} plus Mintz in~\cite{c}). Simultaneous existence of $^{3}\text{H}({^{3}\text{He}},\gamma){^{6}\text{Li}}$
in the ground state and $^{13}\text{C}({^{3}\text{He}},\gamma\alpha){^{12}\text{C}}$
in the ground state in the same experimental setup is indicative of a pre-existing $^{3}\text{H}$ 
structure in $^{12}{\text C}$~\cite{f}. Also $(h,t)$ structure in $A=3$ transfer reactions
is evident for $^{12}{\text C}$ and other neighboring nuclei~\cite{g}.
As such we may take $^{12}{\text C}$ as made up of clusters 2t+2h. In fact MiniBooNE
and NOMAD data both simultaneously taken, be treated as a strong justification of the
wave function given in Eq.(2)

One of the authors (SAA) has aready written several papers where t(and h) appears to be
playing fundamental role in various physical structures (see for example~\cite{h}). It may be 
remarked that the picture offered here may also be used for explaing the quenching of
Gamow-Teller strength obtained in $\text{(p,n)}$ and $(^{3}\text{He},^{3}\text H)$ reactions in nuclei~\cite{i}.

Also a new SU(2) symmetry named as 'nusospin' symmetry where $(h,t)$ form a fundamental representation
has already been suggested~\cite{h}. Just as $(p,n)$ are nucleons in $SU(2)_I$ isospin,$(h,t)$ are 
called 'nusons'
in $SU(2)_A$ nusospin group.

In conclusion, we suggest that miniBooNE group should try at identifying a knocked out 
proton in coincidence with a knocked out helion as per Eq.(3) and Eq.(4). This will enable them
to extract the two strengths simultaneously. This also offers an advantage for antineutrino case where $\bar{\nu_\mu}+p\rightarrow
n+\mu^+$ and $\bar{\nu_\mu}+h\rightarrow t+\mu^+$. Whereas knocked out n is not easily detectable
while the outgoing triton could be easily identified due to its charge. 

One of the author (SAA) would like to thank Joe Grange (MiniBooNE group) for stimulating discussions while he was visiting
our department recently.


\end{document}